\documentclass[letterpaper,12pt]{article}
\usepackage[utf8]{inputenc}
\usepackage{amsmath}
\usepackage{amsfonts}
\usepackage{amsthm}
\usepackage{graphicx}
\usepackage{authblk}
\usepackage{array}
\usepackage{subcaption}
\usepackage{float}
\usepackage{color}
\usepackage{hyperref}
\usepackage{setspace}
\title{Comment concerning Leonardo's rule.}
\author[1]{Sotolongo-Costa, O.}
\author[2]{Gaggero-Sager, L. M.}
\author[3]{Oseguera-Manzanilla, T.}
\author[1]{Díaz-Guerrero, D. S.}
\affil[1]{CInC-Autonomous University of Morelos State, Cuernavaca, Morelos.}
\affil[2]{CIICAp-Autonomous University of Morelos State, Cuernavaca, Morelos.}
\affil[3]{CIDC-Autonomous University of Morelos State, Cuernavaca, Morelos.}
\begin{document}

\maketitle

Leonardo da Vinci's rule concerning tree branching states that ``all the branches of a tree at every stage of its height when put together are equal in thickness 
to the trunk'' \cite{LeonardoNotebook}, i.e., if L is the radius of a branch which ramifies in two branches of radius, say, $L_1$ and $L_2$, then
$L^2\approx L_1^2+L_2^2$. With this obsevation Leonardo establishes a qualitative link between the morphology of ramification and the flow of sap.\newline

In \cite{Eloy2011} it is proposed that Leonardo's rule is a consequence of the self-similarity of the tree trunk and wind-induced stress. In the 
mentioned paper some ad hoc hypothesis, based in fractal theory and fracture theory, were introduced in order to obtain Leonardo's law. 

However, it is 
curious that Leonardo himself proposes a very simple explanation of this rule based on the characteristics of fluid motion. ``When a branch grows, Leonardo 
argues, its thickness will depend on the amount of sap it receives from the one below the branching point. In the tree as a whole, there is a 
constant flow of sap, which rises up through the trunk and divides between the branches flowing through succesive ramifications. Since the total 
quantity of sap carried by the tree is constant, the quantity carried by each branch will be proportional to its cross section, so the total 
cross section at each level will be equal to that of a trunk'' \cite{Capranbook}.

This argument is naive, since the flux of the sap, as viscous liquid, is not proportional to the cross section of the branch, but to the fourth 
power of its radius, if we consider branches and trunk as cylinders.

In our opinion, though Leonardo's rule is not a trivial result from fluid mechanics, a simple and direct approach, based on mass 
conservation,  can be made to explain this observation.

The amount of fluid that goes through a conduit of this type is given by the Poiseuille's Law
\begin{equation}
Q(R)=\frac{\pi\nabla P}{8\eta}R^4.\text{ } Q\text{ is the flux}.
\end{equation}

Now let us add the fact that the xilema of a tree is composed of several conduits with radii distributed according 
to a given law. In each of those conduits the flow obeys Poiseuille's law. Hence the total amount of fluid in a branch 
of width $L$ that contains a distribution of conduits per radius $f(R)$ is 
\begin{equation}
Q_T=\int_0^L Q(R)f(R)dR
\label{QT}
\end{equation}

and if we assume that the main part of  the distribution of conduits per radius, i. e., those that transport most of the sap, 
is of type $f(R)\approx R^{-x}$ (scaling), then the integral gives
\begin{equation}
Q_T\approx L^{5-x}.
\label{qt}
\end{equation}

Hence, in order to get Leonardo's rule we must have that $x=3$. 

If we rely on the assumption that branch structure is 
determined by flow dynamics in a power law distribution of sap conduits, Leonardo's rule can be understood in a very simple way. 
Therefore, let us make the following single hypothesis: the 
distribution of conduits per size (radius) is such that their histogram have a heavy tail.

\begin{figure}[H]
\begin{subfigure}[t]{2.5in}
\includegraphics[width=2in,height=2in]{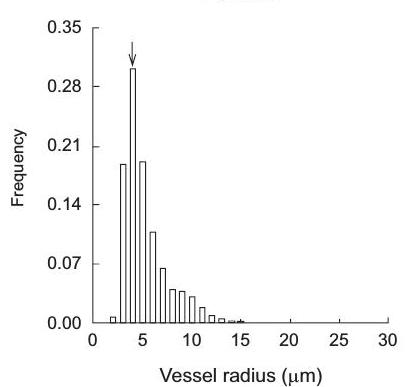}
\caption{Histogram of the conduit radius taken from \cite{Atala2008}. For \textit{betula pendula roth} note the heavy tail of the distribution. 
The tail obeys a scaling law $R^{-x}$ starting from the maximum (pointing arrow).}
\label{histogram}
\end{subfigure}\hfill
\vspace{0.25in}
\begin{subfigure}[t]{2.5in}
\includegraphics[width=2.5in,height=1.95in]{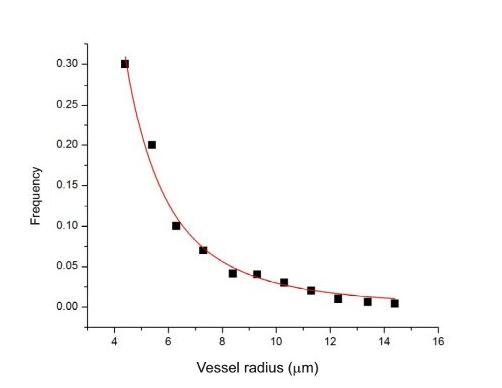}
\caption{A curve for the heavy tail of the histogram. Taking the frequency from \ref{histogram} we adjust the $R^{-x}$ curve that better fits the data in order to obtain 
the value of $x$.}
\label{curve}
\end{subfigure}
\end{figure}

This hypothesis has experimental validation. 
Indeed, in \cite{Atala2008} the \textit{betula pendula roth} xylema distribution is presented. The distributions were measured for different 
parts of the trees. In all cases the distributions are far from the gaussian, exhibiting asymmetry and a heavy tail. 
In this case we are interested only in the tail 
of the distribution since the total flux of the tree is determined by the larger vessels, which carry most of the sap. In figure \ref{histogram} we reproduce one of 
their results, in this case for the branches of the petiole (a slender stem that supports the blade of a foliage leaf). The comparison of the 
data with the measurment gives a correlation coefficient of 0.98.

So, it is possible to take from this distribution the corresponding one for vessels larger than the maximum value, since the main flow occurs through them.
Indeed, in this case the vessels with radius larger than $5$ $\mu$m distribute as shown in figure \ref{curve}. The distribution clearly exhibits 
a fat tail characteristic of Levy distributions. The adjustment of equation \ref{qt} with the data gives $x=3.17$. What gives according to \ref{QT} 
a satisfactory agreement with Leonardo's law.

This result, at least for the case of \textit{betula pendula roth}, is not casual. Other parts if this species were measured and always the xilema 
distribution is such that the self-similarity of the tail is evident. We have no notice of xilema measurements in other trees, though the ubiquity 
of Levy distributions in nature allows us to think that Leonardo's law have a simpler explanation than that proposed in \cite{Eloy2011}.

In our opinion, other ``rules'' by Leonardo could be understood starting from basic laws of physics. Some of our results will be published elsewhere.

\end{document}